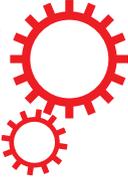

# SCIENTIFIC REPORTS

# Giant field enhancement in high-index dielectric subwavelength particles

Polina Kapitanova[1], Vladimir Ternovski[2], Andrey Miroshnichenko[3], Nikita Pavlov [1], Pavel Belov[1], Yuri Kivshar[1,3] & Michael Tribelsky[2,4]

Besides purely academic interest, giant field enhancement within subwavelength particles at light scattering of a plane electromagnetic wave is important for numerous applications ranging from telecommunications to medicine and biology. In this paper, we experimentally demonstrate the enhancement of the intensity of the magnetic field in a high-index dielectric cylinder at the proximity of the dipolar Mie resonances by more than two orders of magnitude for both the TE and TM polarizations of the incident wave. We present a complete theoretical explanation of the effect and show that the phenomenon is very general – it should be observed for any high-index particles. The results explain the huge enhancement of nonlinear effects observed recently in optics, suggesting a new landscape for all-dielectric nonlinear nanoscale photonics.

Scattering of electromagnetic waves by subwavelength particles underpins important multidisciplinary research, whose applications in physics and chemistry vary from understanding the colors of sky to characterizing colloidal solutions[1]. Many of such effects can be described by the exact solutions of the Maxwell equations[2].

Study of the scattering of light by metallic nanoparticles has attracted a lot of attention due to the excitation of resonant plasmonic modes, which provide a powerful tool for the manipulation and control of light at the nanoscale[3]. One of the main concepts underpinning the field of nanophotonics is large enhancement of electromagnetic fields in the so-called hot-spots of these nanoparticles[3]. Such an enhancement is important for many applications ranging from medicine[4], drug and gene delivery[5], to high-density data recording and storage[6]. However, plasmonic resonances in metal nanoparticles usually are accompanied by large dissipative losses[7].

Therefore, recently a new direction has emerged focused on all-dielectric nanophotonics that makes possible to manipulate optical resonances of dielectric nanoparticles with high refractive index exciting various Mie-type resonant modes selectively[8–12]. Their advantages are attributed to relatively high Q-factors of the excited distinct optical resonances and small mode volumes, providing physical grounds for engineering of new optical functionalities and manufacture of novel metadevices with unique properties[13]. The dielectric particles demonstrate low losses and may exhibit huge concentration of the incident electromagnetic field inside the particles, which has been repeatedly discussed in literature (see, e.g., refs 14–19). However, most experiments in this subfield are related to near-field measurements outside the particles — direct detection of the field inside a subwavelength particle is a difficult (but appealing!) experimental problem.

In the present paper we report a giant enhancement of electromagnetic field in a dielectric cylinder with high refractive index in the proximity of both magnetic and electric dipolar resonances and explain this effect theoretically. The effect is detected experimentally by direct measurements of the field within the cylinder. We observe that the measured intensity of the magnetic field is more than 300 times larger than that of the incident wave, provided the Mie resonant conditions are satisfied. This effect is generic and should be observed for other orders of the Mie-type resonances and other geometries.

[1]ITMO University, St. Petersburg, 197101, Russia. [2]Lomonosov Moscow State University, Moscow, 119991, Russia. [3]Nonlinear Physics Centre, Research School of Physics and Engineering, Australian National University, Canberra, ACT, 2601, Australia. [4]National Research Nuclear University MEPhI (Moscow Engineering Physics Institute), Moscow, 115409, Russia. Polina Kapitanova, Vladimir Ternovski and Andrey Miroshnichenko contributed equally to this work. Correspondence and requests for materials should be addressed to A.M. (email: andrey.miroshnichenko@anu.edu.au)

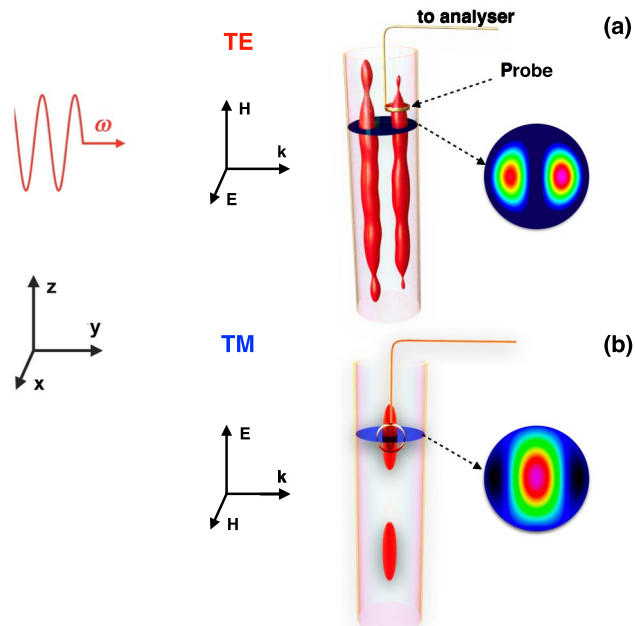

**Figure 1.** Schematic of the experiment. Linearly polarized plane electromagnetic wave with the circular frequency $\omega$ and vector k aligning parallel to the $y$-axis is scattered by a long hollow polystyrene cylinder, whose axis is parallel to the $z$-axis. The cylinder is filled with distilled water known to possess a high refractive index at the microwave frequencies. Magnetic field is measured by a probe placed inside the cylinder. The field profile and probe orientation are shown for two independent polarizations of the incident wave: (**a**) TE and (**b**) TM, respectively. The red balloons schematically represent surfaces of a constant value of $|\mathbf{H}|^2$. See the text for details.

## Methods

**Experiment.** We study the Mie scattering by a dielectric cylinder with a high refractive index $m \equiv \sqrt{\varepsilon} = n + i\kappa$ and low losses ($0 < \mathrm{Im}\,\varepsilon \ll \mathrm{Re}\,\varepsilon$), as shown in Fig. 1. Here $\varepsilon$ stands for the complex permittivity of the cylinder. Similar to the recent experimental studies[20, 21], we simulate a nanowire scattering of a monochromatic plane linearly polarized wave with the visible range frequency by the macroscopic size cylinder scattering of microwave radiation. To this end we employ a hollow polystyrene cylinder filled with distilled water, known to be an excellent material with large dielectric constant and relatively low losses[22–25].

The inner radius of the cylinder is 1.8 cm, the wall thickness is 0.2 cm and the hight is 30 cm. The cylinder is blocked from the bottom side with a polystyrene stopper, whereas from the top side it is open to dip a probe into it.

The cylinder is situated into an anechoic chamber. A rectangular horn antenna (TRIM 0.75–18 GHz; DR) connected to the first transmitting port of a vector network analyser (VNA) Agilent E8362C is used to approximate a plane wave excitation. The antenna is placed at the 2 m distance from the cylinder. To measure the field distribution inside the cylinder, the well known, conventional near-filed scanning technique is employed[26–28]. The experimental setup is shown in Supplementary Information.

To scan the field inside the cylinder, a coaxial cable with the corresponding probe attached to its end is fixed at the arm of the scanner, being connected to the second port of the VNA. During the experiments the probe is placed in the depth of 3 cm below the water level.

The field pattern is studied at the frequency $f = \omega/2\pi$ of the incident wave equal to 1.28 GHz, or 1.3 GHz for the TE (Transverse Electric — vector **E** of the incident wave is perpendicular to the axis of the cylinder) polarization and 1.25 GHz for the TM (Transverse Magnetic — vector **H** of the incident wave is perpendicular to the axis of the cylinder) polarization. It should be stressed, that at the given radius of the cylinder its ratio to the wavelength of the incident wave at the employed frequencies $f = 1.25$; 1.28 and 1.3 GHz is 0.113, 0.115 and 0.117, respectively. Thus, the cylinder is a subwavelength scatterer indeed.

The real and imaginary parts of water permittivity depend on its temperature and the frequency of the incident wave[22, 23]. In our study we use commercially available distilled water. This piece of water may have microimpurities, affecting the permittivity. To know the actual frequency and temperature dependence of the permittivity of the water sample under use in the required frequency and temperature domains we carried out the corresponding experimental study, see Supplementary Information.

To find a compromise between a high value of the real part of the permittivity and its low imaginary part, an appropriate temperature is chosen. For the experiments with the TE polarization we chose the water temperature equals 90 °C. At this temperature the water permittivity at the given frequencies of the electromagnetic radiation (1.28; 1.30 GHz) is $\varepsilon \approx 58.29 + 1.15i$. For the TM polarization measurements the room temperature (25 °C) is employed. At the specified frequency of the incident wave (1.25 GHz) it corresponds to $\varepsilon = 78.05 + 4.89i$.

To prevent the cooling of water during the measurements at the TE polarization the sidewalls of the cylinder are covered with a 6-cm-thick thermal insulating material. Polystyrene and the thermal insulating material, both are, practically, transparent at the GHz frequencies and do not affect the diffraction of the incident wave, see also below *Numerical simulation*. It is important to mention, that, despite the taken measures, the water still cools down through the open top section of the cylinder and the metallic coaxial cable holding the probe inside the cylinder. However, the temperature drop does not exceed 10 °C during the measurement of the entire field profile.

The selected values of the parameters drive the cylinder to the proximity of the dipolar resonance for both polarizations. Electric and magnetic fields within the cylinder are measured by the corresponding probes embedded into the water. Since the electric field enhancement is much smaller than that for the magnetic field, in what follows only results for the magnetic field are discussed.

To measure the magnetic field, we use a small loop as a probe, see Supplementary Information for details. Owing to the problem symmetry at the TE polarization the magnetic field within the cylinder has a single non-zero component aligned along the *z*-axis. To measure it the plane of the probe loop is parallel to the *x*-*y* plane. For the TM polarized excitation wave the magnetic field inside the cylinder has two non-zero components: $H_x$ and $H_y$. To measure them both the probe loop is first aligned parallel to the *y*-*z* plane and and then parallel to the *x*-*z* plane. Next, to find the net field intensity the square of the modula of the two measured components are summed up in each scanning point.

**Numerical simulation.** The simulation of the three-dimensional field distribution in the finite size cylinder are performed with the help of *CST Microwave Studio™* and *Lumerical Solutions* commercial softwares. In these simulations, the water cylinder with radius of 1.8 cm was surrounded by a vacuum and excited by a plane wave, i.e., neither the sidewalls with the thermal insulating material, nor the distortion of the field by the probe are taken into account. The excellent agreement between this numerical simulation and the experimental data (see below) is a convincing evidence that the field perturbations by the probe and the effect of the thermal insulation are negligible indeed.

**Analytical approach.** Then, we employ the well known exact solution of the two-dimensional scattering problem[1] to demonstrate how the observed field enhancement can be described analytically. It makes possible to characterize experimental data and numerical results more specifically. From the theoretical point of view, the problem splits into two parts: (i) study of the scattering fields outside the cylinder, and (ii) study of the internal field excited within the cylinder by the incident plane wave.

Regarding the first problem, it is known that the scattering properties of a high-index dielectric particles are similar to those of the same particle made of a perfect electric conductor (PEC)[1, 17, 29]. However, this is not the case for the inner problem. Specifically, for the PEC the field within the particle vanishes. In contract, for a high-index dielectric the wavelength within the particle decreases, when its refractive index increases. Then, at large enough *n* the size of the particle becomes an integer multiple of the half of the wavelength, no matter how small the geometric size of the particle is. It gives rise to resonances, which, in turn, bring about a giant enhancement of the electromagnetic field within the particle. A detailed analysis of these resonances for a sphere is produced in the recent publication[17]. In what follows, we present an anologous theoretical analysis of the resonant in-field enhancement for a high-index cylinder solving the corresponding two-dimensional Mie scattering problem.

## Results

**Experiment and numerical simulation.** Both experimental data and numerical results are presented in Figs 2 and 3 for the two polarizations, respectively. A great advantage of a microwave range with respect to visible light is that in the former both, the amplitude and the phase of the forward scattered wave may be measured simultaneously. It makes possible the straightforward application of the optical theorem[1, 30] to recover from these measurements the absolute value of the extinction cross section. The extinction cross section obtained in such a manner in our experiments and its comparison with the corresponding quantity calculated in the 3D numerics is presented in Fig. 2(c).

The simulated and measured field profiles agree well with each other, and they exhibit more than 300-fold enhancement of $|\mathbf{H}|^2$ within the cylinder relative to $H_0^2$ [see Fig. 2(a,b,d,e)]. The smaller enhancement of the magnetic field for the TM polarization is explained by the larger value of Im $\varepsilon$ at the room temperature relative to that at 90 °C employed for the TE polarization.

Also, we notice that for the TE polarization in the close vicinity of the dipolar resonance the position of the absolute maximum of the field distribution in a given perpendicular to the cylinder axis section depends on the fine tuning of the frequency of the exciting wave and coincides either with the left or right local maximum, cf. Fig. 2(a,b,d,e). The effect exists already in the 2D case. In the 3D case it is supplemented by the longitudinal modulations caused by the interference of waves, reflected from the top and bottom surfaces of the cylinder. The corresponding patterns are different for the two disconnected "isomagnetic" surface $|H_z|^2 = const$, see the sketch in Fig. 1 and also depend on the frequency of the incident wave. However, detailed discussions of this issue lies beyond the scope of the present paper.

**Theory.** In the present subsection we are interested in the electric (**E**) and magnetic (**H**) fields excited within an infinite circular cylinder by a linearly polarized plane electromagnetic incident wave. They may be written with the help of the two vector harmonics[1]:

$$\mathbf{M}_\ell = k e^{i\ell\phi} \left\{ i\ell \frac{J_\ell(m\rho)}{m\rho}, \, -J_\ell'(m\rho), \, 0 \right\},$$

(1)

$$\mathbf{N}_\ell = k e^{i\ell\phi} \{0, \, 0, \, J_\ell(m\rho)\},$$

(2)

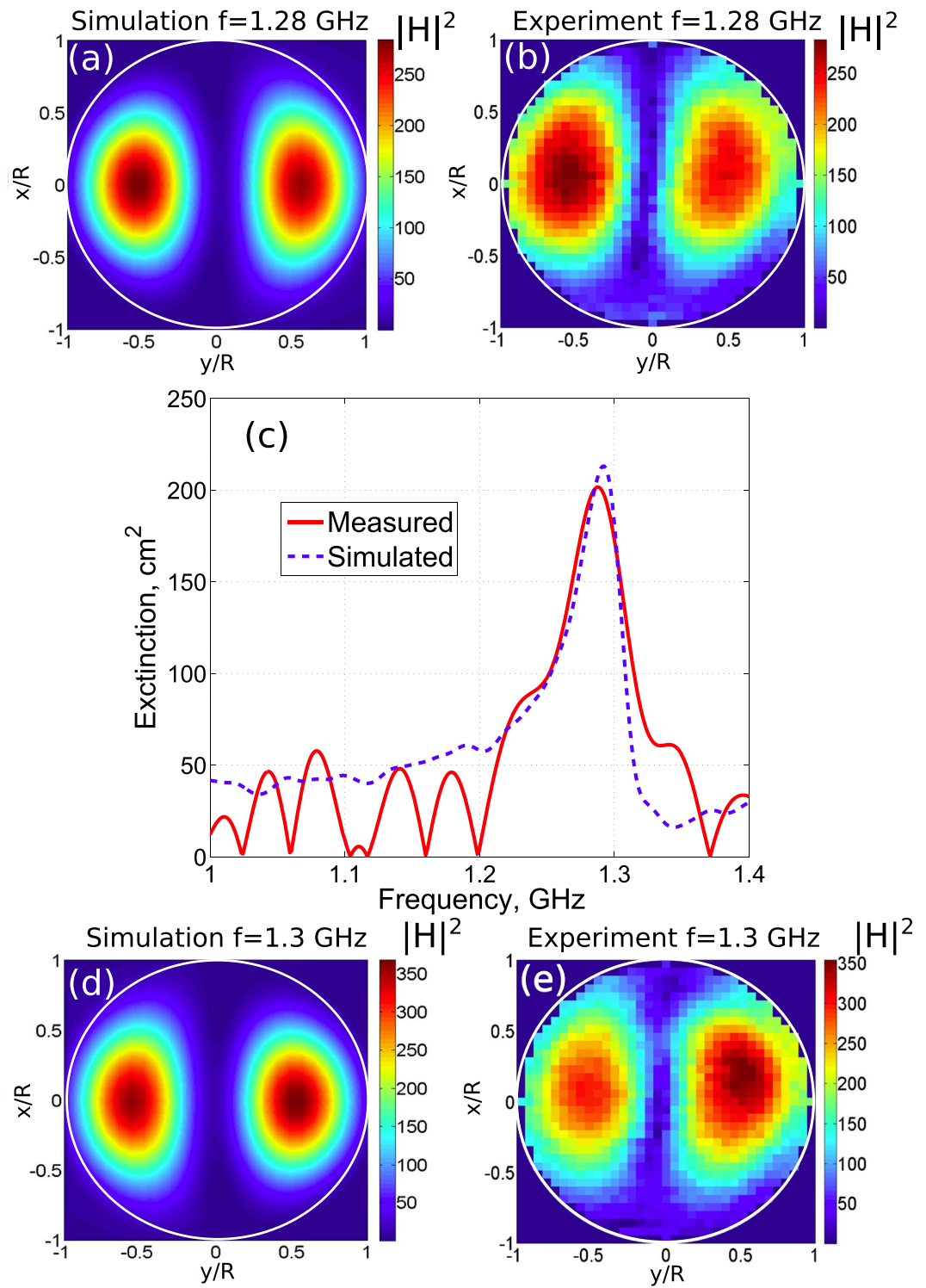

**Figure 2.** Numerically simulated and experimentally measured magnetic field distributions in the cylinder cross section placed at the 3 cm distance from its top surface at the frequency (**a**,**b**) 1.28 GHz and (**d**,**e**) 1.3 GHz for the TE polarized plane wave excitation. The magnetic field is normalized over the incident wave component $H_0$. (**c**) Measured and simulated extinction cross sections of the cylinder, which is filled with distilled water at temperature 90 °C. The incident plane wave propagates along the *y*-axis. The axis of the cylinder is parallel to the *z*-axis.

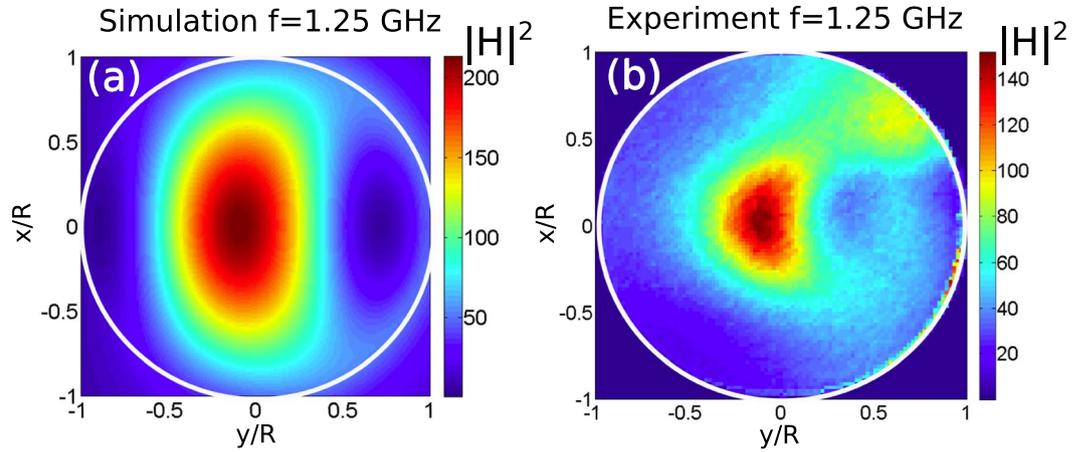

**Figure 3.** The same as that in Fig. 2, but at the frequency 1.25 GHz for the TM polarized plane wave excitation. The cylinder is filled with distilled water kept at the temperature 25 °C.

where $\{F_r, F_\phi, F_z\}$ denote the corresponding components of a vector in the cylindrical coordinate frame, whose $z$-axis is directed along the axis of the cylinder; the prime stands for the derivative with respect to the entire argument of a function; $\rho \equiv rk$ and $k = \omega/c = 2\pi f/c$ stands for the wavenumber of the incident wave in a vacuum (we remind that $m \equiv \sqrt{\varepsilon}$ is the complex refractive index of the cylinder).

In this case, the complex amplitudes of the electric and magnetic fields within the cylinder, normalized over the corresponding quantities of the plane incident waves ($E_0$, $H_0$), can be presented as infinite series of partial modes $\mathbf{E}_\ell$, $\mathbf{H}_\ell$, with $\ell$ varying from minus to plus infinity, and

$$\frac{\mathbf{E}_\ell^{(TE)}}{E_0} = d_\ell \frac{(-i)^{\ell+1}}{k} \mathbf{M}_\ell, \quad \frac{\mathbf{H}_\ell^{(TE)}}{H_0} = -m d_\ell \frac{(-i)^\ell}{k} \mathbf{N}_\ell \qquad (3)$$

$$\frac{\mathbf{E}_\ell^{(TM)}}{E_0} = c_\ell \frac{(-i)^\ell}{k} \mathbf{N}_\ell, \quad \frac{\mathbf{H}_\ell^{(TM)}}{H_0} = m c_\ell \frac{(-i)^{\ell+1}}{k} \mathbf{M}_\ell. \qquad (4)$$

Here the subscripts TE and TM denote the corresponding polarizations. Regarding the coefficients $c_\ell$ and $d_\ell$, they are given by the expressions

$$d_\ell = \frac{2i}{\pi x} \frac{1}{m J_\ell(mx) H_\ell^{(1)'}(x) - H_\ell^{(1)}(x) J_\ell'(mx)}, \qquad (5)$$

$$c_\ell = -\frac{2i}{\pi x} \frac{1}{m J_\ell'(mx) H_\ell^{(1)}(x) - H_\ell^{(1)'}(x) J_\ell(mx)}, \qquad (6)$$

where $x \equiv kR$ is the so-called *size parameter*, $R$ stands for the radius of the cylinder; $J_\ell(z)$ denotes the Bessel function of the first kind. $H_\ell^{(1)}(z) \equiv J_\ell(z) + i Y_\ell(z)$ is the Hankel function of the first kind; $Y_\ell(z)$ is the Bessel function of the second kind, and we have employed the identity

$$J_\ell(x) H_\ell^{(1)'}(x) - J_\ell'(x) H_\ell^{(1)}(x) \equiv \frac{2i}{\pi x}. \qquad (7)$$

Note that $d_{-\ell} \equiv d_\ell$ and $c_{-\ell} \equiv c_\ell$, according to the well known properties of the Bessel functions with integer $\ell$ : $Z_{-\ell}(z) \equiv (-1)^\ell Z_\ell(z)$. Next, at $m = 1$ (i.e., when the optical properties of the cylinder are the same as those of a surrounding vacuum and the cylinder does not scatter the incident light at all) Eqs (5)–(7) yield $c_\ell = d_\ell = 1$. It means that the departure of $|c_\ell|$ and $|d_\ell|$ from unity may be used as a quantitative measure of the resonant field enhancement within the cylinder.

Another important point is that for a given partial mode excited in a high-index cylinder the characteristic relative value of the magnetic field within the scatterer is $|m|$ larger than that for the electric field, see Eqs (1)–(4). The latter is a general property of the problem valid for scatterers of different shapes (cf. the corresponding results for a sphere discussed in ref. 17).

Analysis of Eqs (5) and (6) is quite analogous to that performed in ref. 17. The only difference between the problem in question and the one discussed in ref. 17 is that the Riccati-Bessel functions entering into the Mie solution for a sphere inspected in ref. 17 in our case should be replaced by the Bessel functions. Bearing it in mind, the results obtained in ref. 17 maybe be adopted for the problem under consideration straightforwardly. In particular, at large $n$ and small $\kappa$ the resonant values of $x$ at the leading approximation in $1/n$ are defined by the conditions

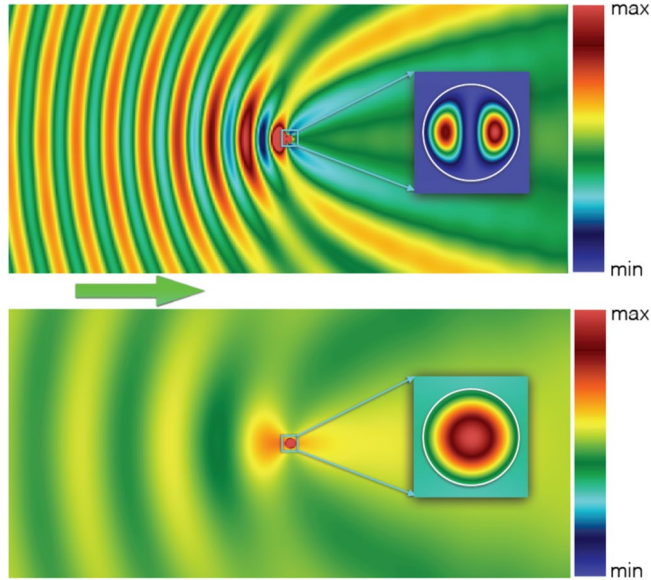

**Figure 4.** Images obtained with the help of the exact Mie solution of the two-dimensional problem. Spatial distribution of the intensity of the magnetic field ($|\mathbf{H}|^2$), normalized over the corresponding quantity in the incident wave ($H_0^2$), at light scattering by an infinite dielectric cylinder with radius $R = 1.8$ cm and permittivity $\varepsilon = 58.29 + 1.15i$. The green arrow indicates the direction of propagation of the plane TE polarized incident wave. The upper panel corresponds to the electric dipolar resonance at $f = 1.31$ GHz (cf. Fig. 2); the lower one shows the off-resonance case at at $f = 0.5$ GHz. The field within the cylinder is shown in the inserts. The white circle in the inserts designates the cylinder boundary. *The color bars have the three different scales*: (i) for the field outside the cylinder min = 0.5, max = 2; (ii) for the field in the upper panel insert min = 0, max = 522 (iii) for the field in the lower panel insert min = 0, max = 3.2. Note strong spatial modulations of the backward scattered field outside the cylinder and a deep suppression of the forward scattering at the resonant frequency.

$$J_\ell(nx_{TE}^{(0)}) = 0, \quad J_\ell'(nx_{TM}^{(0)}) = 0, \tag{8}$$

see Eqs (5) and (6). The resonance line shapes for the coefficients $|d_n|^2$, $|c_n|^2$ have the Lorentzian profiles, and their explicit form is presented in Supplementary Information.

To get an idea of applications of the general theoretical results discussed above to the performed experiments, we present the explicit expression for the sum of two partial modes $\mathbf{H}_{\pm\ell}^{(TE)} \equiv \mathbf{H}_{-\ell}^{(TE)} + \mathbf{H}_{\ell}^{(TE)}$. The only nonzero component of $\mathbf{H}_{\pm\ell}^{(TE)}$ is $(H_{\pm\ell}^{(TE)})_z$, see Eq. (3). Then, simple calculations give rise to the following formula:

$$\frac{(H_{\pm\ell}^{(TE)})_z}{H_0} = -2(-i)^\ell m d_\ell J_\ell(m\rho) \cos(\ell\phi). \tag{9}$$

Specifically, in the case corresponding to the experiments: $\ell = 1$, $\varepsilon = 58 + 1.15i$, (i.e., $n = 7.64$, $\kappa = 0.075$) and the resonance occurs in the vicinity of $x = 0.5$. In this case, according to Eq. (8), $x_{TE}^{(0)} \approx 0.5018$, while the corresponding maximum of $|d_\ell|^2$ is achieved at $x \approx 0.4923$ and equals 6.80. Then, as follows from Eq. (9), the profile of $|(H_{\pm\ell}^{(TE)})_z/H_0|^2$ has two maxima situated at $\phi = 0$, $\phi = \pi$ and $r/R \approx 0.49$. At these two local maxima, we have $\max\{|(H_{\pm\ell}^{(TE)})_z/H_0|^2\} \approx 536.6$ (cf. Fig. 4).

This field profile and the maximal values of $\{|(H_{\pm\ell}^{(TE)})_z/H_0|^2\}$ are very close to the ones obtained in the experiments and the 3D simulation, see Fig. 2. The TM case is treated analogously.

## Discussion

We have presented a direct experimental evidence of the dramatic enhancement of electromagnetic fields at the Mie resonances of high-index dielectric particles. We have observed more than 300-fold enhancement of the magnetic field measured inside a dielectric cylinder for the scattering of plane TE polarized electromagnetic waves, and about 150-fold enhancement, for the TM polarized electromagnetic waves. We have demonstrated numerically, with the help of three-dimensional simulations, that this enhancement occurs due to the selective excitation of the Mie-type dipolar modes, and we also confirm this result analytically with the help of the two-dimensional Mie scattering theory.

We believe these results would modify the established viewpoint on the wave scattering by subwavelength particles with high refractive indices, which may play a role of effective subwavelength resonators of a novel type, enhancing the fields by several orders of magnitude. Being scaled to the optical wavelengths, our results explain the nature of the recent experimental observations of the multifold enhancement of nonlinear effects at light scattering by Si-based nanostructures[31, 32]. In a broad perspective they may initiate creation of a new landscape for all-dielectric nonlinear nanophotonics and stimulate further study in this important and challenging field.

## References

1. Bohren, C. F. & Huffman, D. R. *Absorption and Scattering of Light by Small Particles* (Wiley: New York, 1983).
2. Mie, G. Beiträge zur optik trüber medien, speziell kolloidaler metallösungen. *Annalen der Physik* **330**, 377–445, doi:10.1002/(ISSN)1521-3889 (1908).
3. Novotny, L. & Hecht, B. *Principles of Nano-optics* (Cambridge University Press, New York, 2006).
4. Anderson, R. R. & Parrish, J. A. Selective photothermolysis: Presice microsurgery by selective absorption of pulsed radiation. *Science* **220**, 524–527, doi:10.1126/science.6836297 (1983).
5. Han, G., Ghosh, P., De, M. & Rotello, V. M. Drug and gene delivery using gold nanoparticles. *Nano Biotechnology* **3**, 40–45, doi:10.1016/j.addr.2008.03.016 (2007).
6. Pan, L. & Bogy, D. B. Data storage: Heat-assisted magnetic recording. *Nature Photonics* **3**, 189–190, doi:10.1038/nphoton.2009.40 (2009).
7. Klimov, V. *Nanoplasmonics* (Pan Stanford, 2014).
8. Kuznetsov, A. I., Miroshnichenko, A. E., Fu, Y. H., Zhang, J. B. & Lukyanchuk, B. Magnetic light. *Scientific Reports* **2**, 492, doi:10.1038/srep00492 (2012).
9. Evlyukhin, A. B. *et al.* Demonstration of magnetic dipole resonances of dielectric nanospheres in the visible region. *Nano Letters* **12**, 3749–3755, doi:10.1021/nl301594s (2012).
10. Staude, I. *et al.* Tailoring directional scattering through magnetic and electric resonances in subwavelength silicon nanodisks. *ACS Nano* **7**, 7824–7832, doi:10.1021/nn402736f (2013).
11. Tribelsky, M. I., Geffrin, J.-M., Litman, A., Eyraud, C. & Moreno, F. Small dielectric spheres with high refractive index as new multifunctional elements for optical devices. *Scientific Reports* **5**, 12288, doi:10.1038/srep12288 (2015).
12. Kuznetsov, A. I., Miroshnichenko, A. E., Brongersma, M., Kivshar, Y. S. & Luk'yanchuk, B. S. Optically resonant dielectric nanostructures. *Science* **354**, aag2472 (2016).
13. Zheludev, N. & Kivshar, Yu. S. From metamaterials to metadevices. *Nature Materials* **11**, 917–924, doi:10.1038/nmat3431 (2012).
14. Cao, L. *et al.* Engineering light absorption in semiconductor nanowire devices. *Nature Materials* **8**, 643–647, doi:10.1038/nmat2477 (2009).
15. García-Etxarri, A. *et al.* Strong magnetic response of submicron Silicon particles in the infrared. *Opt. Expr.* **19**, 4815–4826, doi:10.1364/OE.19.004815 (2011).
16. Traviss, D. J., Schmidt, M. K., Aizpurua, J. & Muskens, O. L. Antenna resonances in low aspect ratio semiconductor nanowires. *Opt. Expr.* **23**, 22771–22787, doi:10.1364/OE.23.022771 (2015).
17. Tribelsky, M. I. & Miroshnichenko, A. E. Giant in-particle field concentration and Fano resonances at light scattering by high-refractive-index particles. *Phys. Rev. A* **93**, 053837, doi:10.1103/PhysRevA.93.053837 (2016).
18. van de Groep, J., Coenen, T., Mann, S. A. & Polman, A. Direct imaging of hybridized eigenmodes in coupled silicon nanoparticles. *Optica* **3**, 93–99, doi:10.1364/OPTICA.3.000093 (2016).
19. Coenen, T., van de Groep, J. & Polman, A. Resonant Modes of Single Silicon Nanocavities Excited by Electron Irradiation. *ACS Nano* **7**, 1689–1698, doi:10.1021/nn3056862 (2013).
20. Rybin, M. V., Filonov, D. S., Belov, P. A., Kivshar, Yu. S. & Limonov, M. F. Switching from visibility to invisibility via Fano resonances: Theory and experiment. *Scientific Reports* **5**, 8774, doi:10.1038/srep08774 (2015).
21. Rybin, M. V. *et al.* Phase diagram for the transition from photonic crystals to dielectric metamaterials. *Nature Communications* **6**, 10102, doi:10.1038/ncomms10102 (2015).
22. Ellison, W. J. Permittivity of pure water, at standard atmospheric pressure, over the frequency range 0 ÷ 25 THz and the temperature range 0 ÷ 100 °C. *J. Phys. Chem. Ref. Data* **36**, 1–18, doi:10.1063/1.2360986 (2007).
23. Kaatze, U. Complex permittivity of water as function of frequency and temperature. *J. of Chem. and Engineering Data* **34**, 371–374, doi:10.1021/je00058a001 (1989).
24. Andryieuski, A., Kuznetsova, S. M., Zhukovsky, S. V., Kivshar, Yu. S. & Lavrinenko, A. V. Water: Promising opportunities for tunable all-dielectric electromagnetic metamaterials. *Scientific Reports* **5**, 13535, doi:10.1038/srep13535 (2015).
25. Odit, M., Kapitanova, P., Andryieuski, A., Belov, P. & Lavrinenko, A. V. Experimental demonstration of water based tunable metasurface. *Appl. Phys. Lett.* **109**, 011901, doi:10.1063/1.4955272 (2016).
26. Dutta, S. K. *et al.* Imaging microwave electric fields using a near-field scanning microwave microscope. *Appl. Phys. Lett.* **74**, 156–158, doi:10.1063/1.123137 (1999).
27. Whiteside, H. & King, R. The loop antenna as a probe. *IEEE Trans. Antennas & Propag.* **12**, 291–297 (1964).
28. Kapitanova, P. V. *et al.* Photonic spin Hall effect in hyperbolic metamaterials for polarization-controlled routing of subwavelength modes. *Nature Communications* **5**, 3226, doi:10.1038/ncomms4226 (2014).
29. Landau, L. D., Lifshitz, E. M. & Pitaevskii, L. P. *Electrodynamics of Continuous Media* (Pergamon Press, Oxford 1984).
30. Newton, R. G. Optical theorem and beyond. *American Journal of Physics* **44**, 639–642, doi:10.1119/1.10324 (1976).
31. Yang, Y. *et al.* Nonlinear Fano-Resonant Dielectric Metasurfaces. *Nano Lett.* **15**, 7388–7393, doi:10.1021/acs.nanolett.5b02802 (2015).
32. Shorokhov, A. S. *et al.* Multifold Enhancement of Third-Harmonic Generation in Dielectric Nanoparticles Driven by Magnetic Fano Resonances. *Nano Lett.* **16**, 4857–4861, doi:10.1021/acs.nanolett.6b01249 (2016).

## Acknowledgements

The authors acknowledge the assistance of M. Odit in measurements of the distilled water permittivity. The experimental study of the field enhancement was supported by the Russian Science Foundation (project N14-12-00897). The numerical simulations presented in this work were supported by the Government of the Russian Federation (projects No. GZ 3.561.2014/K and GZ 2014/190) and the Russian Foundation for Basic Research (projects No. 14-02-31761), a grant of the President of Russian Federation (MD-7841.2015.2), and the Australian Research Council. P.K. acknowledges a scholarship of the President of the Russian Federation. V.T. acknowledges the support from CR Research, Inc. Hong Kong.

## Author Contributions

P.K., V.T., and M.T. developed the initial idea and prepared the experimental setup. P.K. performed the numerical simulations in CST Microwave Studio and supervised the experimental study. V.T. performed analytical analysis and optimized parameters of the structure for the maximal field enhancement. N.P. conducted the experiment and performed the near-field scanning. A. M. visualized analytical results, performed modal analysis, and direct numerical simulations. Y.K., P.B. and M.T. supervised the whole project. All authors participated in the writing of the manuscript and contributed equally.

## Additional Information

**Supplementary information** accompanies this paper at doi:10.1038/s41598-017-00724-5

**Competing Interests:** The authors declare that they have no competing interests.

**Publisher's note:** Springer Nature remains neutral with regard to jurisdictional claims in published maps and institutional affiliations.